\documentclass[10pt,conference]{IEEEtran}
\IEEEoverridecommandlockouts

\usepackage{cite}
\usepackage[utf8]{inputenc}
\usepackage{quantikz}
\usepackage{hyperref}
\usepackage{amsmath}
\usepackage{amsfonts}
\usepackage{braket}
\usepackage{xcolor}

\newsavebox{\boxA}
\newsavebox{\boxB}
\newsavebox{\boxC}

\usepackage[caption=false]{subfig}

\usepackage{geometry}
\usepackage{color}

\hypersetup{hidelinks,colorlinks=true,breaklinks=true,urlcolor=blue,allcolors=blue,pdftitle={Title},pdfauthor={Author}}

\def\BibTeX{{\rm B\kern-.05em{\sc i\kern-.025em b}\kern-.08em
    T\kern-.1667em\lower.7ex\hbox{E}\kern-.125emX}}

\begin{document}

\title{Transferability of optimal QAOA parameters between random graphs}

\author{\IEEEauthorblockN{Alexey Galda\IEEEauthorrefmark{1},
Xiaoyuan Liu\IEEEauthorrefmark{3},
Danylo Lykov\IEEEauthorrefmark{2},
Yuri Alexeev\IEEEauthorrefmark{2},
and Ilya Safro\IEEEauthorrefmark{3}}
\IEEEauthorblockA{\IEEEauthorrefmark{1} James Franck Institute, University of Chicago, Chicago, IL 60637 USA}
\IEEEauthorblockA{\IEEEauthorrefmark{2}Computational Science Division, Argonne National Laboratory, Lemont, IL 60439 USA}
\IEEEauthorblockA{\IEEEauthorrefmark{3}Department of Computer and Information Sciences, University of Delaware, 
Newark, DE 19716, USA}
Email: \IEEEauthorrefmark{1}agalda@uchicago.edu}

\maketitle{}

\begin{abstract}
The Quantum approximate optimization algorithm (QAOA) is one of the most promising candidates for achieving quantum advantage through quantum-enhanced combinatorial optimization. In a typical QAOA setup, a set of quantum circuit parameters is optimized to prepare a quantum state used to find the optimal solution of a combinatorial optimization problem. Several empirical observations about optimal parameter concentration effects for special QAOA MaxCut problem instances have been made in recent literature, however, a rigorous study of the subject is still lacking. We show that convergence of the optimal QAOA parameters around specific values and, consequently, successful transferability of parameters between different QAOA instances can be explained and predicted based on the local properties of the graphs, specifically the types of subgraphs (lightcones) from which the graphs are composed. We apply this approach to random regular and general random graphs. For example, we demonstrate how optimized parameters calculated for a 6-node random graph can be successfully used without modification as nearly optimal parameters for a 64-node random graph, with less than 1\% reduction in approximation ratio as a result. This work presents a pathway to identifying classes of combinatorial optimization instances for which such variational quantum algorithms as QAOA can be substantially accelerated.
\end{abstract}

\begin{IEEEkeywords}
quantum computing, quantum optimization, quantum approximate optimization algorithm
\end{IEEEkeywords}

\section{Introduction}
Quantum computing seeks to exploit the quantum mechanical concepts of entanglement and superposition to perform a computation that is significantly faster and more efficient than what can be achieved using the most powerful supercomputers available today. Demonstrating quantum advantage with optimization algorithms~\cite{Alexeev2021} is poised to have a broad impact on science and humanity by allowing to solve problems of a global scale, including energy, materials discovery, and environmental challenges. Variational quantum algorithms are considered as primary candidates for such tasks and consist of parameterized quantum circuits with parameters updated in a classical computation. The quantum approximate optimization algorithm (QAOA)~\cite{Hogg2000quantumsearch, Hogg2000, farhi2014quantum, hadfield2017quantum} is a variational algorithm for solving classical combinatorial optimization problems. In the domain of optimization on graphs, it is most often used to solve such NP-hard problems as MaxCut~\cite{farhi2014quantum}, community detection~\cite{shaydulin2019network} and partitioning~\cite{ushijima2021multilevel} by mapping them onto a classical spin-glass model (also known as the Ising model) and minimizing the corresponding energy, which in itself is NP-hard.

In this work, we demonstrate that by analyzing the distributions of subgraphs from two QAOA MaxCut instance graphs, it is possible to predict how close the optimized QAOA parameters for one instance are to the optimal QAOA parameters for another. The measure of transferability of optimized parameters between MaxCut QAOA instances on two graphs can be expressed through the value of the approximation ratio, which is defined as the ratio of the energy of the corresponding QAOA circuit, evaluated with the optimized parameters $\gamma, \beta$, divided by the energy of the optimal MaxCut solution for the graph. 
While the optimal solution is not known in general, for relatively small instances (graphs with up to 64 nodes, considered in this paper) it can be found using classical algorithms. We first focus our attention on random regular graphs of arbitrary degree and reveal that transferability of optimized parameters MaxCut QAOA between two graphs is directly determined by the transferability between all possible permutations of pairs of individual subgraphs. The relevant subgraphs of these graphs are defined by the QAOA quantum circuit depth parameter $p$. In this work, we focus on the case $p = 1$, however, our approach can be extended to larger values of $p$. Higher values of $p$ lead to an increasing number of subgraphs to be considered, however the general idea of the approach remains the same. This question is beyond the scope of this paper and will be addressed in our future work.

Based on the analysis of mutual transferability of optimized QAOA parameters between all relevant for computing the MaxCut cost function subgraphs of random regular graphs, we reveal good transferability \emph{within} the classes of odd- and even-regular random graphs of arbitrary size.
We also show that transferability is poor \emph{between} the classes of even- and odd-regular random graphs, in both directions, based on the poor transferability of the optimized QAOA parameters between the subgraphs of the corresponding graphs. We then consider the most general case of arbitrary random graphs, construct the transferability map between all possible subgraphs of such graphs, with an upper limit of node connectivity ${d_\mathrm{max} = 6}$, and use it to demonstrate that in order to find optimized parameters for a MaxCut QAOA instance on a large 64-node random graph, under specific conditions, it is possible to re-use the optimized parameters from a random graph of a much smaller size, ${N = 6}$, with only a 0.8\% reduction in the approximation ratio.

This paper is structured as follows. In Section~\ref{sec:QAOA}, we present the relevant background material on QAOA and tensor network simulation techniques relevant to this work. In Section~\ref{sec:transfer}, we consider optimized QAOA parameter transferability properties between all possible subgraphs of random regular graphs of degree up to ${d_\mathrm{max} = 8}$. We then extend the consideration to arbitrary random graphs, bounded by the maximum degree of connectivity ${d_\mathrm{max} = 6}$, and demonstrate the power of the proposed approach for constructing three pairs of graphs, with 6 and 64-node nodes in each, such that the optimized QAOA parameters found for the smaller graphs can be successfully used for the larger ones. Finally, in Section~\ref{sec:conclusions}, we conclude with a summary of our results and an outlook.

\section{QAOA}\label{sec:QAOA}
The Quantum Approximate Optimization Algorithm is a hybrid quantum-classical algorithm that combines a parameterized quantum evolution with a classical outer-loop optimizer to approximately solve binary optimization problems~\cite{farhi2014quantum,hadfield2019quantum}. QAOA consists of $p$ layers of pairs of alternating operators (also known as a circuit depth), with each additional layer increasing the quality of the solution, assuming perfect noiseless execution of the corresponding quantum circuit. With quantum error correction not currently supported by modern quantum processors, practical implementations of QAOA are limited to ${p \leq 3}$ due to noise and limited coherence of quantum devices imposing strict limitations on the circuit depth. Motivated by the practical relevance of results, we focus on the case ${p = 1}$ in this paper.

\subsection{QAOA background}\label{subsec:theory}

Consider a combinatorial problem defined on a space of binary strings of length $N$ that has $m$ clauses. Each clause is a constraint satisfied by some assignment of the bit string. The objective function can be written as ${C(z) = \sum_{\alpha=1}^m C_{\alpha}(z)}$, where $z = z_1z_2\cdots z_N$ is the bit string, and ${C_{\alpha}(z) = 1}$ if $z$ satisfies the clause $\alpha$, and 0 otherwise. QAOA maps the combinatorial optimization problem onto a $2^N$ dimensional Hilbert space with computational basis vectors $\ket{z}$, and encodes $C(z)$ as an operator $C$ diagonal in the computational basis.

At each call to the quantum computer, a trial state is prepared by applying a sequence of alternating quantum operators 
\newcommand{\gb}{{\vec{\beta}, \vec{\gamma}}}

\begin{equation} \label{eq:QAOAstate}
    \ket{\gb}_p := U_B(\beta_p)U_C(\gamma_p)\ldots U_B(\beta_1)U_C(\gamma_1)\ket{s}\,,
\end{equation}
where $U_C(\gamma) = e^{-i\gamma C}$ is the phase operator, $U_B(\beta)=e^{-i\beta B}$ is the mixing operator, with $B$ defined as the operator of all single-bit $\sigma^x$ operators, $B = \sum_{j=1}^N \sigma_j^x$, and $\ket{s}$ is some easy-to-prepare initial state, usually taken to be the uniform superposition product state. The parameterized quantum circuit (\ref{eq:QAOAstate}) is called the \emph{QAOA ansatz}. We refer to the number of alternating operator pairs~$p$ as the \textit{QAOA depth}. The selected parameters $\vec{\beta},\vec{\gamma}$ are said to define a \emph{schedule}, analogous to a similar choice  in quantum annealing.

Preparation of the state (\ref{eq:QAOAstate}) is followed by a measurement in the computational basis. The output of repeated state preparation and measurement may be used by a classical outer-loop algorithm to select the schedule~$\vec{\beta},\vec{\gamma}$. We consider optimizing the expectation value of the objective function

\[\langle C\rangle_p = \bra{\vec{\beta}, \vec{\gamma}}_pC\ket{\vec{\beta}, \vec{\gamma}}_p \,,
\]
as originally proposed in~\cite{farhi2014quantum}. The output of the overall procedure is the best bit string $z$ found for the given combinatorial optimization problem. We emphasize that the task of finding good QAOA parameters is challenging in general, e.g. due to encountering such challenges as barren plateaus. Acceleration of the optimal parameters search for a given QAOA depth $p$ is the focus of many approaches aimed at demonstrating the quantum advantage. Examples include such methods as warm- and multi-start  optimization~\cite{egger2020warmstarting,shaydulin2019multistart}, problem decomposition~\cite{shaydulin2019hybrid}, and instance structure analysis~\cite{shaydulin2020classical} and parameter learning~\cite{khairy2020learning}.

\subsection{MaxCut}\label{subsec:maxcut}

For the purpose of studying transferability of optimized QAOA parameters, we consider the MaxCut combinatorial optimization problem. Given an unweighted undirected simple graph ${G = (V,E)}$, the goal of the MaxCut problem is to find a partition of the graph's vertices into two complementary sets, such that the number of edges between the two sets is maximized. To encode the problem in the QAOA setting, the input is a graph with ${|V| = N}$ vertices, and ${|E| = m}$ edges, and the goal is to find a bit string $z$ that maximizes \begin{eqnarray} C = \sum_{jk\in E} C_{jk}, \label{eq:maxcut_cost} \end{eqnarray}
where \begin{eqnarray*}
C_{jk} = \frac{1}{2}(-\sigma_j^z \sigma_k^z + 1).\end{eqnarray*}

It has been shown in \cite{farhi2014quantum} that on a 3-regular graph, QAOA with $p=1$ produces a solution with an approximation ratio of at least 0.6924.

\subsection{Tensor network QAOA simulator}\label{subsec:sim}

There are two general approaches for classical simulation of QAOA quantum circuits: state vector and tensor network simulators. In this work, we perform all simulations of QAOA quantum circuits using the QTensor tensor network simulator~\cite{lykov2020tensor}, a large-scale quantum circuit simulator with step-dependent parallelization.

In contrast to state vector simulators, which store the full state vector of size $2^N$, tensor network simulators do not have the notion of evolution of a state vector, but rather view the whole calculation as a tensor network contraction task.
Numerical simulations for this type of simulators can be executed in two ways:
computation of probability amplitudes, and evaluation of an expectation value of some observable. These simulations correspond to calculating elements of the output state $\ket{\phi} = \hat U \ket \psi$, and the value of $\braket{\phi| \hat R | \phi} = \braket{\psi | \hat U^\dagger \hat R \hat U| \psi}$, respectively. Each of these expressions has a tensor network analog, where the operator $\hat U$ is represented by a tensor network that is composed of elementary gates acting on subsets of qubits. For more details on how a quantum circuit is converted to a tensor network, see \cite{Schutski_2020, lykov2020tensor}.

When simulating an expectation value of some observable operator $\hat R$ that acts on a small subset of qubits, most of the constituents of the $\hat U$ operator cancel out. This is known as the lightcone simplification and introduces a major improvement in the simulation cost, especially for circuits with low connectivity and depth.

When applied to the MaxCut QAOA problem, the $\hat R$ operator is a sum of smaller terms, as shown in Eq. \ref{eq:maxcut_cost}.
The expectation value of the cost for the graph $G$ and QAOA depth $p$ is then

\begin{align*}
\braket{C}_p(\gb) &= \braket{\gb| C | \gb} 
\\
&= \braket{\gb| \sum_{jk\in E}\frac{1}{2}(1-Z_jZ_k)|\gb}
\\
&= \frac{|E|}{2} - \frac{1}{2}\sum_{jk\in E}\braket{\gb|Z_jZ_k|\gb}
\\
&\equiv \frac{|E|}{2} - \frac{1}{2}\sum_{jk\in E}e_{jk}(\gb),
\end{align*}
where $e_{jk}$ is an individual edge contribution to the total cost function.
Note that the observable in the definition of $e_{jk}$ is local to only two qubits, therefore most of the gates in the circuit that generates the state $\ket \gb$ cancels out. The circuit after the cancellation is equivalent to calculating $Z_jZ_k$ on a subgraph $S$ of the original graph $G$. 
These subgraphs can be obtained by taking only the edges that are incident from vertices at a distance $p-1$ from the vertices $j$ and $k$.
The full calculation of $E_G(\gb)$ requires evaluation of $|E|$ tensor networks, each representing the value $e_{jk}(\gb)$ for $jk \in E$.

After the tensor network has been created, the simulator has to determine the best way to contract it, which can be done in multiple ways. QTensor uses the bucket elimination approach, which contracts one index of the tensor network at a time. The order in which the indices are contracted determines the largest tensor size, which, in turn, is the main contribution to the memory and time cost for the simulation. The number of dimensions in the largest tensor is commonly referred to as contraction width. The size of the largest tensor is therefore determined as $S = 2^w$ where $w$ is the contraction width.
QTensor is able to simulate tensor networks with $w<30$ in seconds using several GB of RAM.

\begin{figure}
    \centering
    \includegraphics[width=\linewidth]{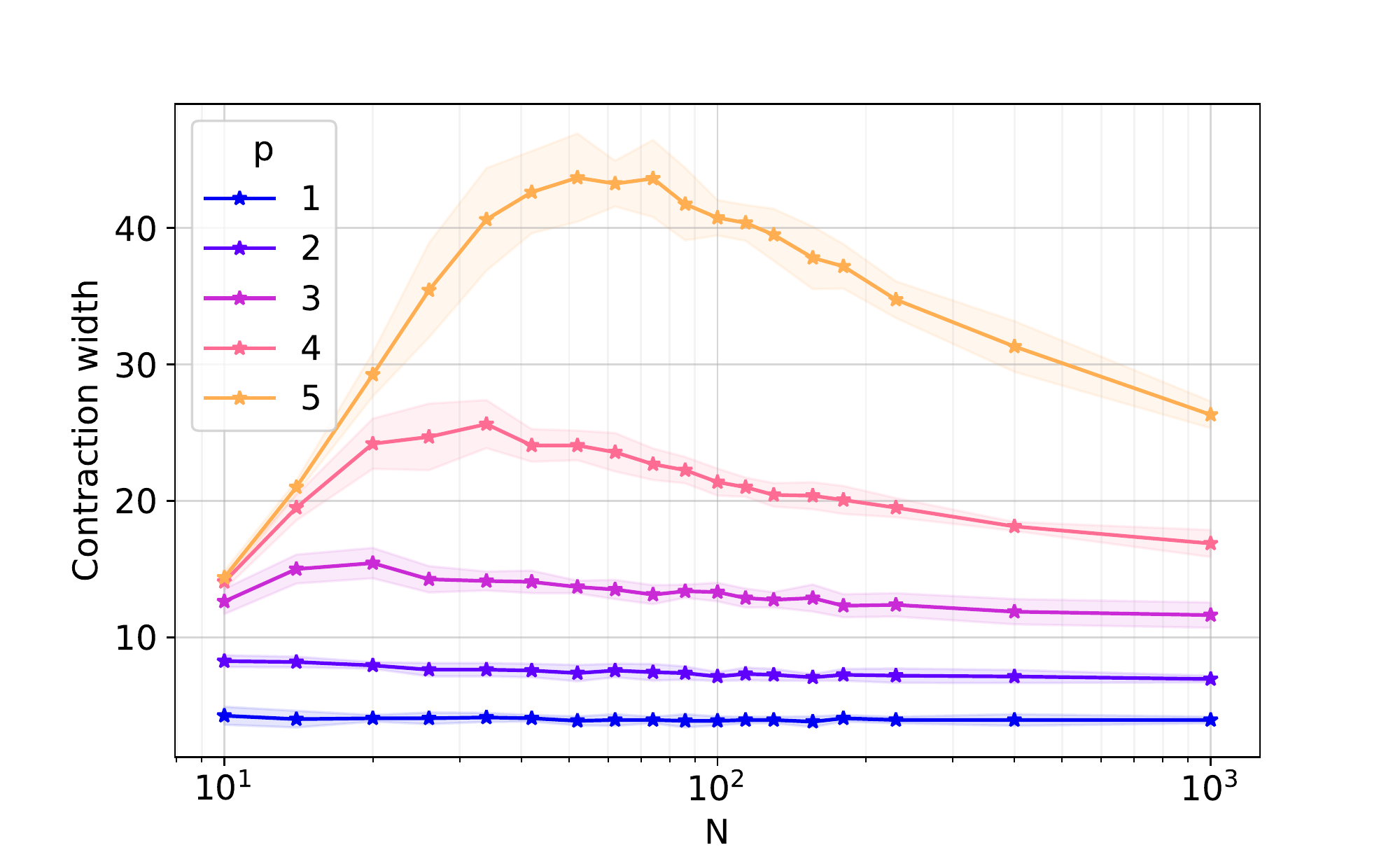}
    \caption{Maximum contraction width for simulating energy using the QTensor simulator. 
    The x-axis shows the size of a random 3-regular graph used to generate MaxCut QAOA circuits.
    The shaded region shows the standard deviation over 16 random graphs for each size.
    }
    \label{fig:simcost}
\end{figure}

\begin{figure}
    \centering
    \includegraphics[width=\linewidth]{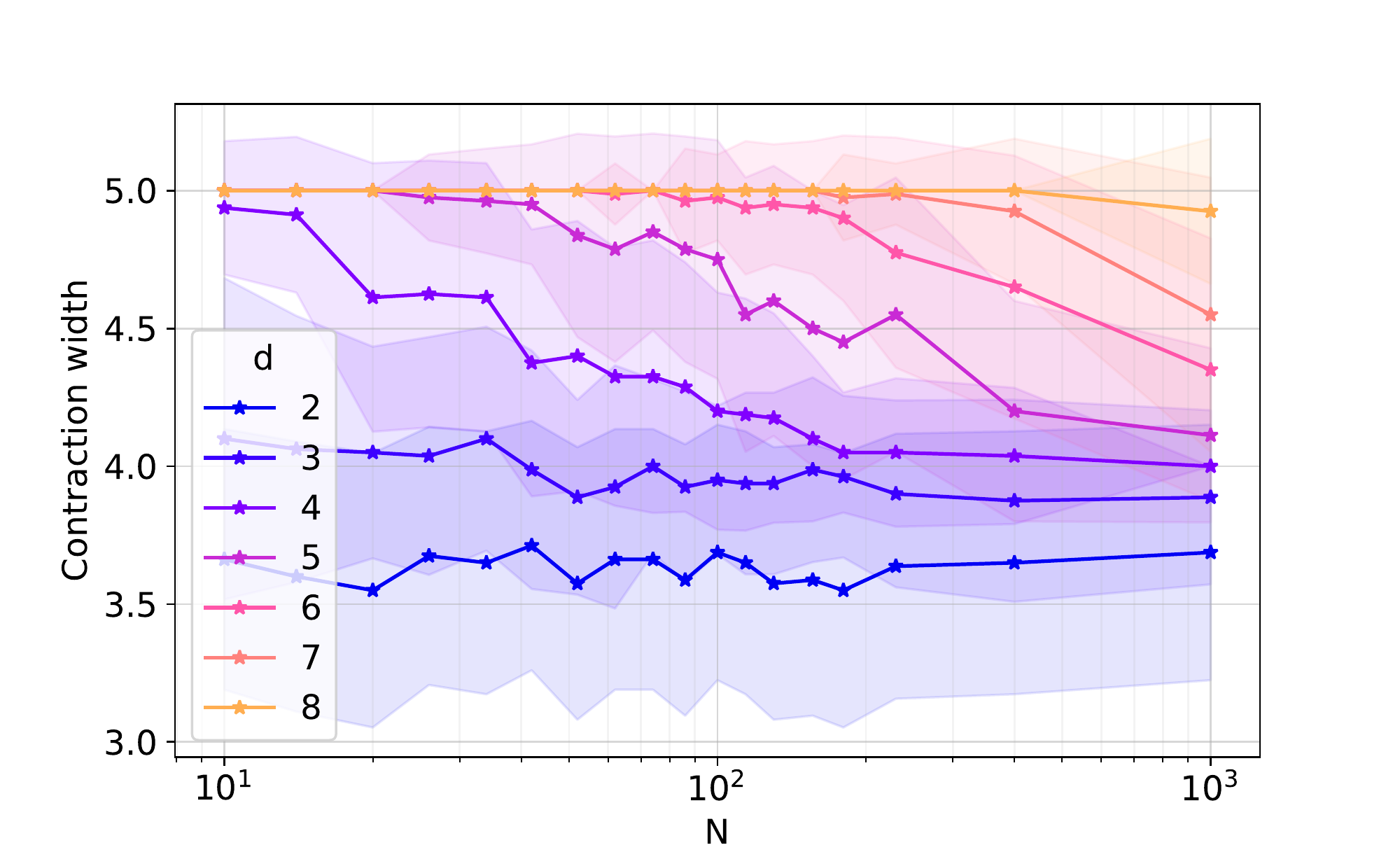}
    \caption{Maximum contraction width for simulating energy using the QTensor simulator. 
    The x-axis shows the size of a random $d$-regular graph used to generate MaxCut QAOA $p=1$ circuits.
    The shaded region shows the standard deviation over 80 random graphs for each size.
    }
    \label{fig:simcost_vs_d}
\end{figure}

In studying the complexity of MaxCut simulation, we focus on three main parameters: degree $d$, size $N$ of the problem graph $G$, and the QAOA depth $p$.
We first fix $d=3$, corresponding to 3-regular random graphs, and study $p$-dependence of simulation cost,
and then look at $d$-dependence for fixed $p=1$.

The tensor network simulations show that for $p<5$, one can simulate energy for any size of a graph, with linear scaling in time. Fig.~\ref{fig:simcost}
shows the dependence of contraction width as a function of graph size $N$ for different QAOA depths $p$. The experiments were done using QTensor with the ordering algorithm \texttt{rgreedy\_0.01\_10}.
The non-monotonic behavior of the simulation complexity with respect to the graph size is a result of the structure of random 3-regular graphs. As size $N$ grows, the probability of large loops in the graph increases, and subgraphs that generate tensor networks for $e_jk$ become more tree-like. This results in a simpler tensor network structure and a smaller contraction cost.
Since the tractability of simulation of energy extends far beyond $p=1$ discussed in this paper, there is a room for further research on this topic.

The dependence of simulation cost at fixed $p=1$ is shown in Fig.~\ref{fig:simcost_vs_d}. The subgraphs for $p=1$ only include edges incident from nodes $jk$ for each $e_jk$. Therefore, the maximum contraction width is $w=5$ for any $N$, and it slowly decreases as more tree-like subgraphs occur in the structure of the graph.

To perform the parameter optimization, QTensor uses automatic differentiation with respect to the gate parameters. Automatic differentiation uses the chain rule to propagate the gradients over the calculation graph, which allows calculating first-order derivatives with respect to any input parameters in one shot. This approach is widely used in the field of machine learning as the backpropagation algorithm.

To optimize the parameters of the quantum circuit, we use \texttt{RMSProp}, a common machine learning algorithm which is an extension of the gradient descent technique. 

\subsection{Classical MaxCut solver}

Calculating the approximation ratio for a particular MaxCut problem instance, requires the optimal solution of the combinatorial optimization problem. This problem is known to be NP-hard, and classical solvers require exponential time to converge. For our experiments, we use the Gurobi solver \cite{gurobi} with the default configuration parameters, running the solver until it converges to the optimal solution.

\section{Parameter transferability}\label{sec:transfer}

Solving a QAOA instance calls for two types of executions of quantum circuits, iterative optimization of the QAOA parameters, and the final sampling from the output state prepared with the those parameters. While the latter is known to be impossible to simulate efficiently for large enough instances using classical hardware instead of a quantum processor~\cite{farhi2014quantum}, the iterative energy calculation for the QAOA circuit during the classical optimization loop can be efficiently performed using tensor network simulators for instances of a wide range of sizes~\cite{lykov2020tensor}, as described in the previous Section. This is achieved by implementing considerable simplifications in how the expectation value of the problem Hamiltonian is calculated by employing a mathematical reformulation based on the notion of the reverse causal cone introduced in the seminal QAOA paper~\cite{farhi2014quantum}. Moreover, in some instances, the entire search of the optimal parameters for a particular QAOA instance can be circumvented by reusing the optimized parameters from a different `related' instance, e.g. for which the optimal parameters are concentrated in the same region. 

\emph{Optimizing QAOA parameters for a relatively small graph, called donor, and using them to prepare the QAOA state that maximizes the expectation value $\langle C\rangle_p$ for the same problem on a larger graph, called acceptor, is what we define as successful optimal parameter transferability, or just transferability of parameters, for brevity.} The transferred parameters can be either used directly without change, as implemented in this paper, or as a `warm start' for further optimization. In either case, the high computational cost of optimizing the QAOA parameters, which grows rapidly as the QAOA depth $p$ and the problem size are increased, can be significantly reduced. This approach presents a new direction for dramatically reducing the overall runtime of QAOA.

\begin{figure*}[h!]
    \centering
    \includegraphics[width=\linewidth]{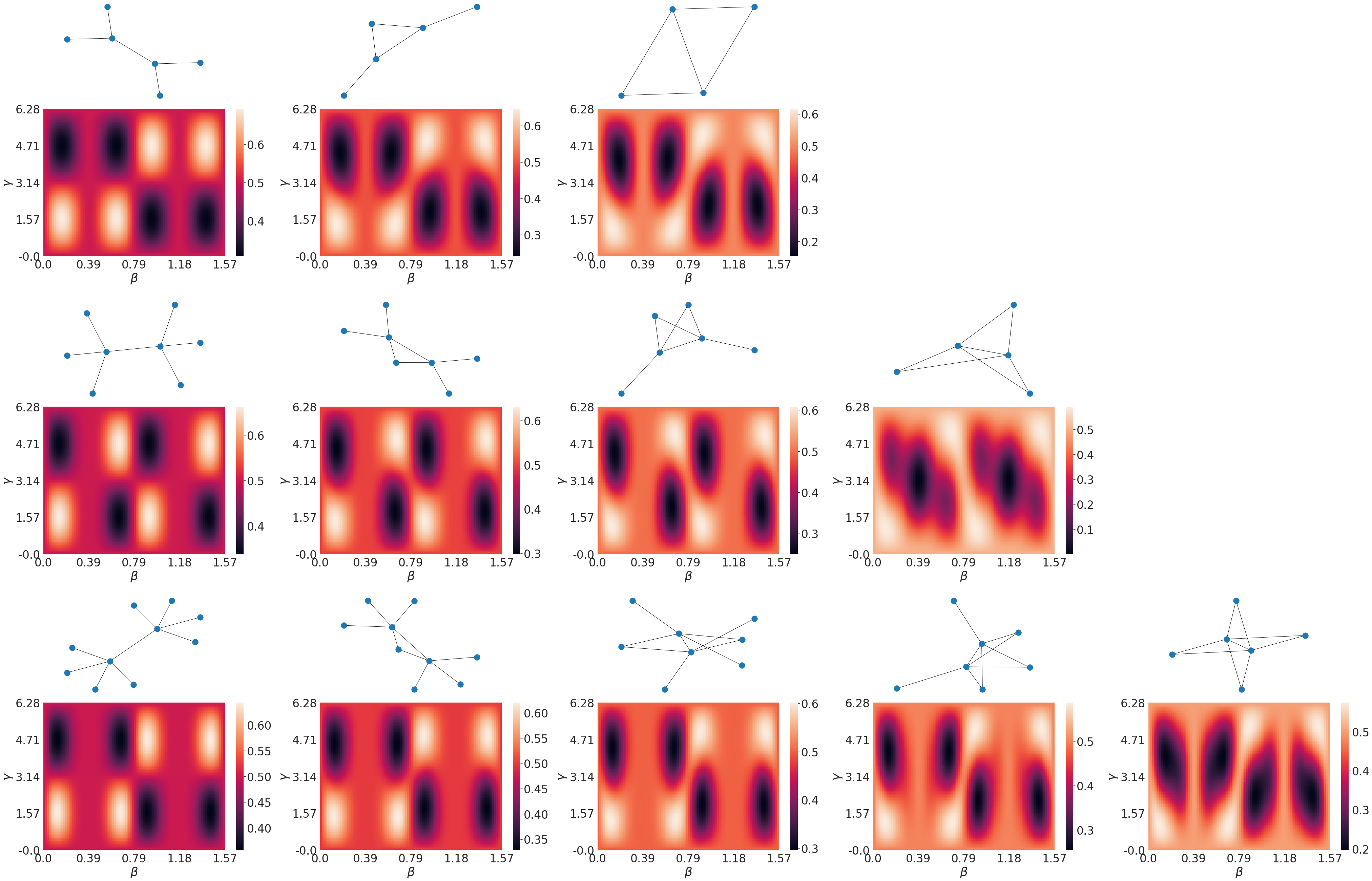}
    \caption{Landscapes of energy contributions for individual subgraphs of 3- (top row), 4- (middle row), and 5-regular (bottom row) random graphs, as a function of QAOA parameters $\beta$ and $\gamma$. All subgraphs of 3- and 5- regular graphs have maxima located in the relative vicinity from one another. Subgraphs of 4-regular graphs also have closely positioned maxima between themselves, however only half of them match with the maxima of subgraphs of odd-regular random graphs.}
    \label{fig:landscapes}
\end{figure*}

Optimal QAOA parameter concentration effects have been reported in the past for several special cases, mainly focusing on random 3-regular graphs~\cite{brandao2018fixed, streif2020training, akshay2021parameter}. Brandao et al.~\cite{brandao2018fixed}, observed that the optimized QAOA parameters for the MaxCut problem obtained for a 3-regular graph are also nearly optimal for all other 3-regular graphs. In particular, it was noted that in the limit of large $N$, where $N$ is the number of nodes, the fraction of tree graphs asymptotically approaches 1. We note that, for example, in the sparse Erd\"os--R\'enyi  graphs, the trees are observed in short distance neighborhoods with very high probability~\cite{newman2018networks}. As a result, in this limit, the objective function is the same for all 3-regular graphs, up to order $1/N$. 

The central question of this manuscript is determining under what conditions the optimized QAOA parameters for one graph also maximize the QAOA objective function for another graph. Because the expectation value of the QAOA objective function is fully determined by the corresponding subgraphs of the instance graph, to study transferability of parameters between graphs, we study the transferability between their subgraphs.

\subsection{Subgraph transferability analysis}\label{subsec:subgraphs}

It was shown in the seminal QAOA paper~\cite{farhi2014quantum} that the expectation value of the QAOA objective function, $\langle C\rangle_p$, can be evaluated as a sum over contributions from subgraphs of the original graph, provided its degree is bounded and diameter is larger than $2p$ (otherwise, the subgraphs cover the entire graph itself). The contributing subgraphs can be constructed by iterating over all edges of the original graph and selecting only the nodes that are $p$ edges away from the edge. Through this process, any graph can be deconstructed into a set of subgraphs for a given $p$, and only those subgraphs contribute to $\langle C\rangle_p$, as also discussed in Section~\ref{sec:QAOA}.

We begin by analyzing the case of MaxCut instances on 3-regular random  graphs for QAOA circuit depth $p = 1$, which have three possible subgraphs~\cite{farhi2014quantum, brandao2018fixed}. Fig.~\ref{fig:landscapes} (top row) shows the landscapes of energy contributions from these subgraphs, evaluated for a range of $\gamma$ and $\beta$ parameters. It is apparent that all maxima are located in approximate vicinity of each other. As a result, the parameters optimized for either of the three graphs will also be near-optimal for the other two. Because any random 3-regular graph can be decomposed into these three subgraphs, for QAOA with $p = 1$, this guarantees that optimized QAOA parameters can be successfully transferred between any two 3-regular random graphs, which is in full agreement with~\cite{brandao2018fixed}. The same effect is observed for subgraphs of 4-regular, see Fig.~\ref{fig:landscapes} (middle row). The optimized parameters are mutually transferable between all four possible subgraphs of 4-regular graphs. Notice, however, that the locations of exactly a half of all maxima for the subgraphs of 4-regular graphs do not match with those for 3-regular graphs. This means that one cannot expect good transferability of optimized parameters across MaxCut QAOA instances for 3- and 4-regular random graphs.
Focusing now on all five possible subgraphs of 5-regular graphs, Fig.~\ref{fig:landscapes} (bottom row), we notice that, again, good parameter transferability is expected between all instances of 5-regular random graphs. Moreover, the locations of the maxima match well with those for 3-regular graphs, indicating good transferability across 3- and 5-regular random graphs.

By considering energy contribution landscaped of subgraphs of random regular graphs for $p=1$ we make the following three conjectures:
\begin{enumerate}
    \item Optimized parameters can be successfully transferred between any random $d$-regular graphs, $d \in \mathbb{N}$.
    \item Optimized parameters can be successfully transferred from any a random $d_1$-regular graph to any random $d_2$-regular graph, assuming that $d_1$ and $d_2$ are either both odd or both even.
    \item Optimized parameters cannot be successfully transferred between $d_1$- and $d_2$-regular graphs, if $d_1$ is odd and $d_2$ is even, or vice versa.
\end{enumerate}
A rigorous study of the question of transferability between random regular graphs is necessary to confirm the above conjectures, and is outside of the scope of this paper. This question, together with the generalization to $p > 1$ and the method of rescaling the optimized parameters for improved transferability between $d_1$- and $d_2$- regular random graphs (for odd or even $d_1$ and $d_2$), with will be considered in a separate paper.

\begin{figure}[b!]
    \centering
    \includegraphics[width=180pt]{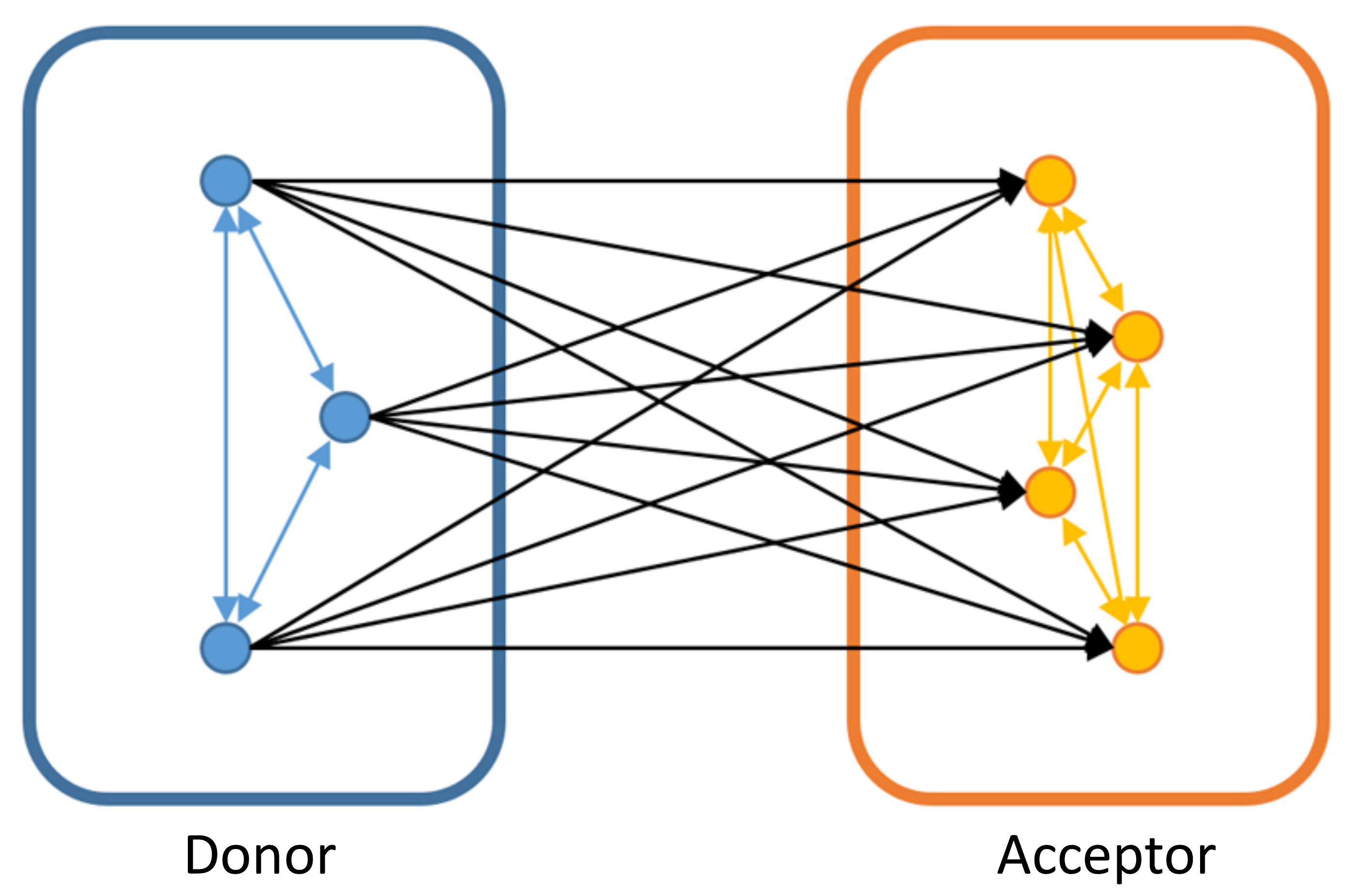}
    \caption{Schematic representation of the sufficient condition for successful transferability of optimal QAOA parameters from a donor to an acceptor graph. Each graph, represented by a stadium, consists of a number of subgraphs, represented by circles. The arrows indicate the directions in which optimal parameters need to be transferable between individual subgraphs in order to guarantee transferability from the donor to the acceptor graph.}
    \label{fig:subgraphs}
\end{figure}

Mutual transferability of optimized parameters between all possible subgraphs of 3-regular graphs is what guarantees that optimized parameters found for some QAOA MaxCut instance on a 3-regular graph are also nearly optimal for any other 3-regular graph. In this work, we provide a significant extension of this result and present the following three conditions as the sufficient conditions for optimal parameter transferability for general donor-acceptor pairs of random graphs (see a schematic example in Fig~\ref{fig:subgraphs}):
\begin{itemize}
    \item optimal QAOA parameters are mutually transferable between all subgraphs of the donor graph;
    \item optimal QAOA parameters are mutually transferable between all subgraphs of the acceptor graph;
    \item optimal QAOA parameters are transferable from every subgraph of the donor graph to every subgraph of the acceptor graph.
\end{itemize}

\begin{figure*}[h!]
    \centering
    \includegraphics[width=\columnwidth*2]{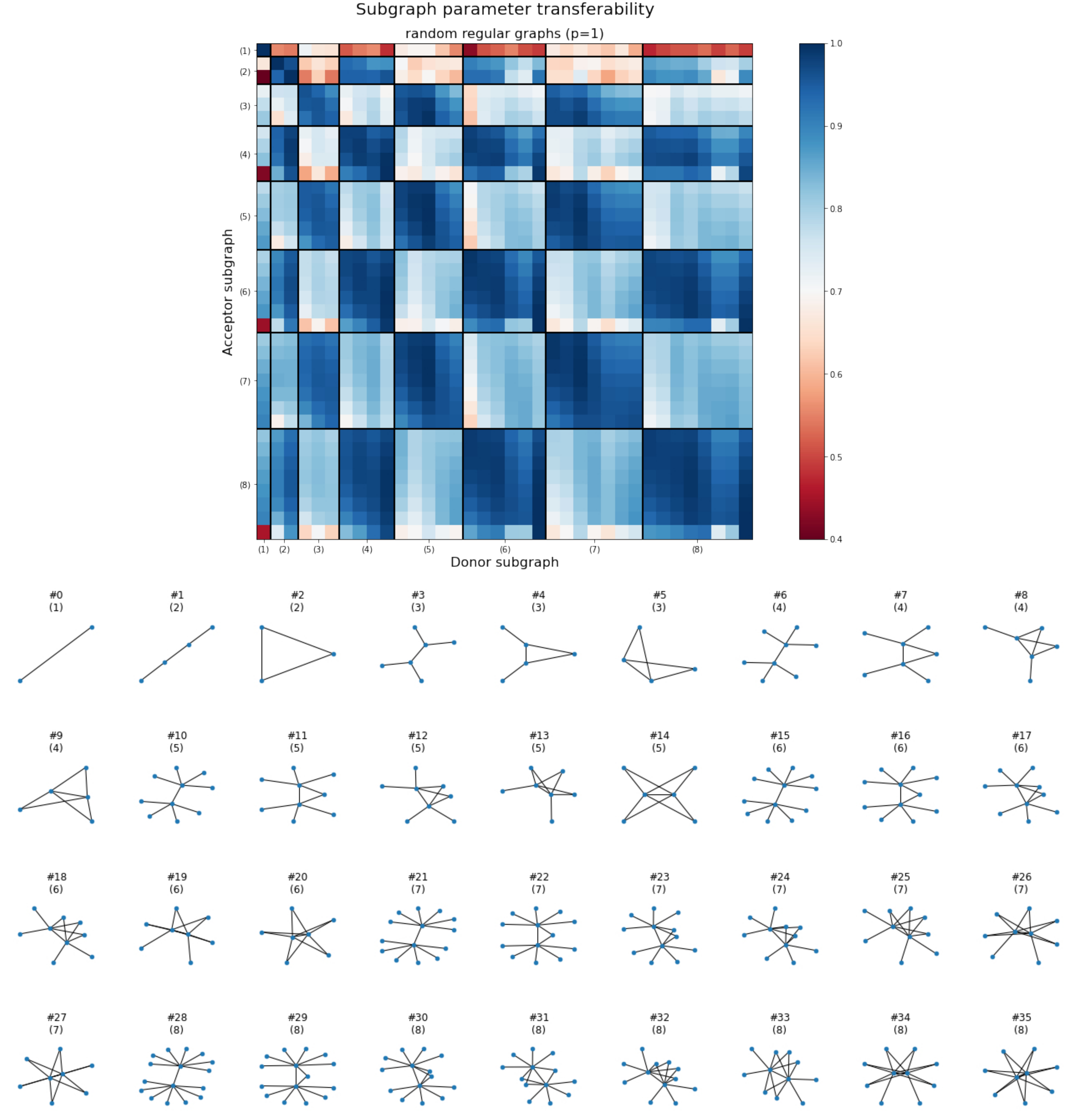}
    \caption{Transferability map between all subgraphs of random regular graphs with maximum node degree $d_\mathrm{max} = 8$, for QAOA depth $p = 1$. High (blue) and low (red) values represent good and bad transferability, correspondingly. Good transferability among even-regular and odd-regular random graphs, and poor transferability across even- and odd-regular graphs, in both directions, is observed.}
    \label{fig:heatmap1}
\end{figure*}

\begin{figure*}[h!]
    \centering
    \includegraphics[width=400pt]{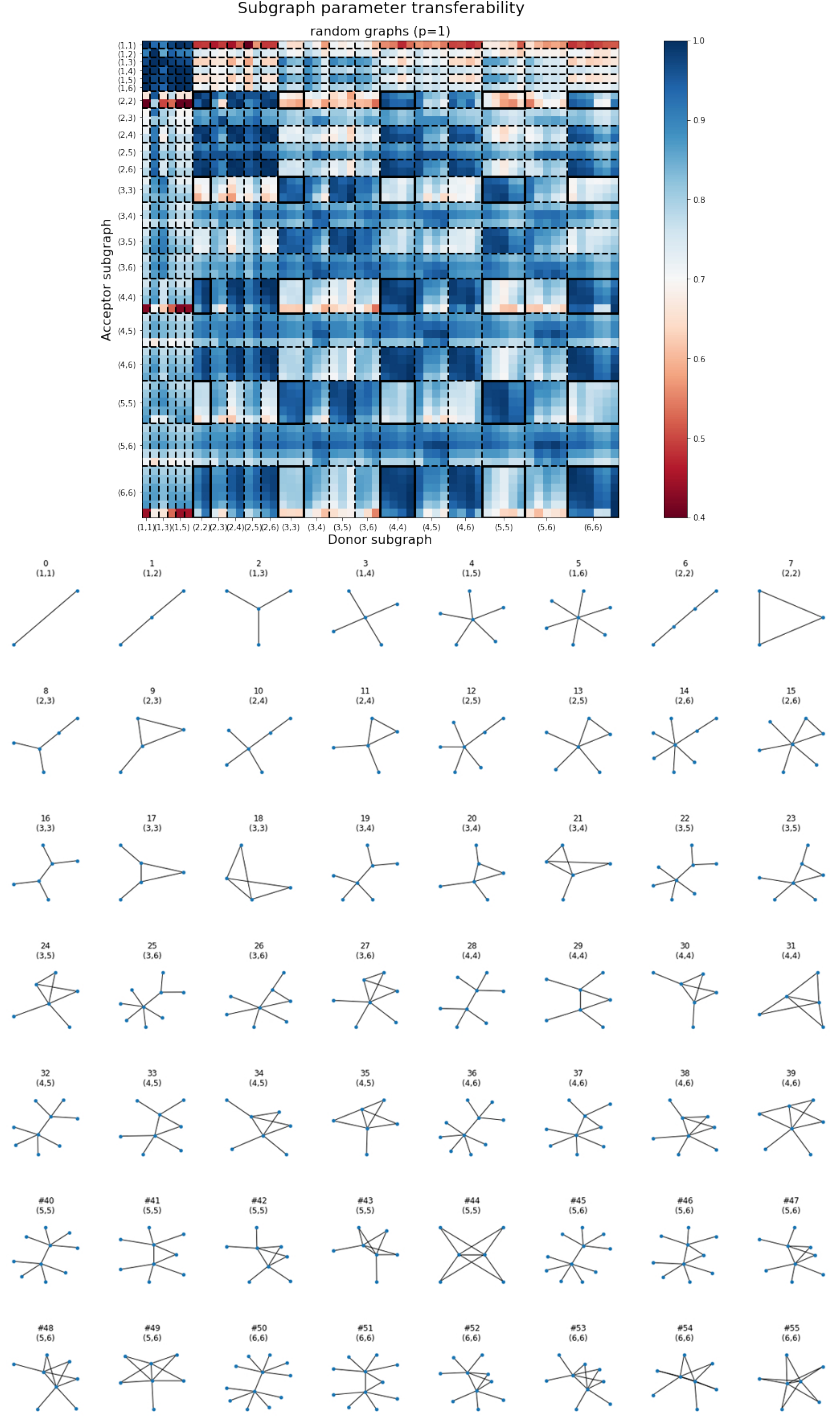}
    \caption{Transferability map between all subgraphs of random graphs with maximum node degree $d_\mathrm{max} = 6$, for QAOA depth $p = 1$. Subgraphs are visually separated by dashed lines into groups of subgraphs with the same degrees of the nodes forming the central edge. Solid black rectangles correspond to optimized parameter transferability between subgraphs of random regular graphs (Fig.~\ref{fig:heatmap1}).}
    \label{fig:heatmap2}
\end{figure*}

For the case of parameter transferability between 3-regular graphs of arbitrary size, the above three conditions are automatically satisfied based on a single fact that optimal parameters are mutually transferable between all three possible subgraphs of 3-regular graphs, as discussed above. To test numerically the three conjectures postulated above regarding transferability of optimized parameters between different random regular graphs, we evaluate the subgraph transferability map between all possible subgraphs of $d$-regular graphs, $d \leq 8$, see Fig.~\ref{fig:heatmap1}. The top panel shows the colormap of parameter transferability coefficients between all possible pairs of subgraphs of $d$-regular graphs ($d \leq 8$, 35 subgraphs total). Each axis is split into groups of $d$ subgraphs of $d$-regular graphs, and the color values in each cell represent the transferability coefficient computed for the corresponding directional pair of subgraphs, defined as follows. For every donor subgraph, from which the optimized $\gamma\,, \beta$ parameters are being transferred, we performed numerical optimization with 200 steps, repeated 20 times with random initial points. Each of the obtained $(\gamma, \beta)$ pairs was then used to evaluate the QAOA energy contribution of the central edge of the corresponding acceptor subgraph. The transferability coefficient was then calculated as the average energy contribution of the acceptor subgraph, evaluated using each of the 20 optimized parameters from the donor subgraph, divided by the maximum energy contribution of the acceptor subgraph found by the same optimization procedure. All considered subgraphs are shown on the bottom panel of Fig.~\ref{fig:heatmap1}. Note that parameter transferability is a directional property between (sub)graphs, and good transferability from (sub)graph A to (sub)graph B does not guarantee good transferability from B to A. This general fact can be easily understood by considering two graphs with commensurate energy landscapes, for which every energy maximum corresponding to graph A also falls onto the energy maximum for graph B, but some of the energy maxima for graph B do not coincide with those of graph A.

The regular pattern of alternating clusters of high and low transferability coefficients in Fig.~\ref{fig:heatmap1} illustrates that the parameter transferability effect extends from 3-regular graphs to the entire family of odd-regular graphs, as well as to even-regular graphs (Conjectures 1 and 2), with poor transferability between the two classes (Conjecture 3). For example, the established result for 3-regular graphs is reflected at the intersection of columns and rows with the label `(3)' for both donor and acceptor subgraphs. The fact that all cells in the 3x3 block in Fig.~\ref{fig:heatmap1}, corresponding to parameter transfer between subgraphs of 3-regular graphs, have high values, representing high mutual transferability, explains the observation of optimal QAOA parameter transferability between arbitrary 3-regular graphs~\cite{brandao2018fixed}.

\subsection{General random graph transferability}\label{subsec:general}

Having considered optimal MaxCut QAOA parameter transferability between random regular graphs, we now focus on general random graphs. Subgraphs of an arbitrary random graph differ from subgraphs of random regular graphs in that the two nodes connected by the central edge can have a different number of connected edges, making the set of subgraphs of general random graphs much more diverse. The upper panel of Fig.~\ref{fig:heatmap2} shows the transferability map between all possible subgraphs of random graphs with node degrees $d \leq 6$, a total of 50 subgraphs, presented in the lower panel. The transferability map can serve as a lookup table for determining whether optimized QAOA parameters are transferable between any two graphs.

Fig.~\ref{fig:heatmap2} reveals another important fact about parameter transferability between subgraphs of general random graphs. Subgraphs labeled as $(i, j)$, where $i$ and $j$ represent the degrees of the two central nodes of the subgraph, are in general transferable to any other subgraph $(k, l)$, provided that all $\{i, j, k, l\}$ are either odd or even. This result is a generalization of the transferability result for odd- and even-regular graphs described above. However, Fig.~\ref{fig:heatmap2} shows that there exist a number of pairs of subgraphs with mixed degrees (not only even or odd) that also transfer well to other mixed degree subgraphs, e.g. $\mathrm{subgraph}\,\#20\,(3, 4) \to \mathrm{subgraph}\,\#34\,(4, 5)$. The map of subgraph transferability provides a unique tool for identifying smaller donor subgraphs, the optimized QAOA parameters for which are also nearly optimal parameters for the original graph. It can also be used to define the likelihood of parameter transferability between two graphs based on their subgraphs. Below we demonstrate how it can explain the optimized QAOA parameter transferability between seemingly unrelated different 6- and 64-node graphs.

\subsection{Parameter transferability examples}

\begin{figure}[h!]
    \centering
    \includegraphics[width=\columnwidth]{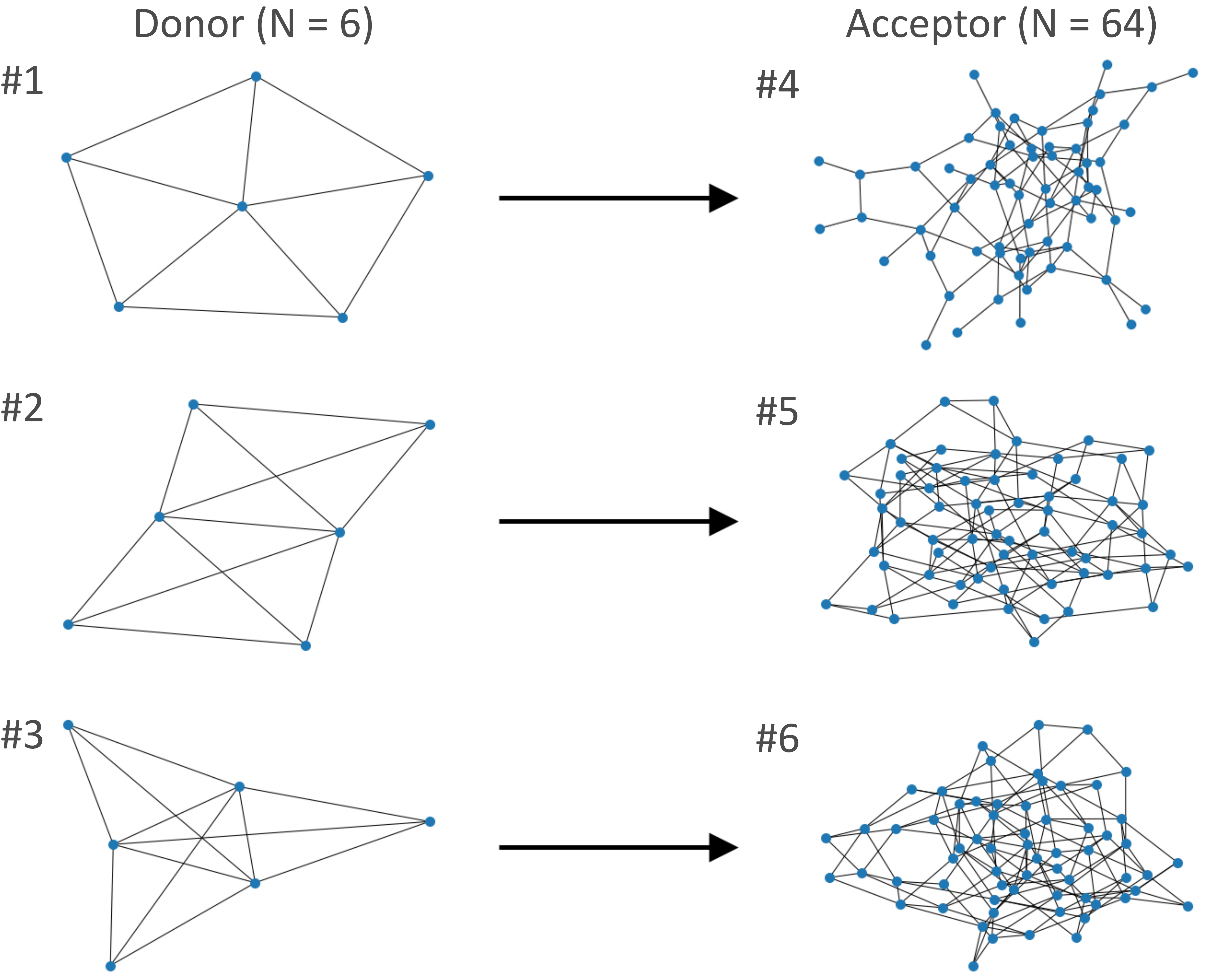}
    \caption{Demonstration of optimized parameter transferability between $N=6$ donor and $N=64$ acceptor random graphs. Using optimized parameters from the donor graph for the acceptor leads to the reduction in approximation ratio of 0.8\%, 2.0\%, and 2.1\% for the three examples, top to bottom, compared to optimizing the parameters for the acceptor graph directly, for $p = 1$.}
    \label{fig:example}
\end{figure}

We will now demonstrate that the parameter transferability map from Fig.~{\ref{fig:heatmap1}} can be used to find small-$N$ donor graphs from which the optimized QAOA parameters can be successfully transferred to a MaxCut QAOA instance on a much larger acceptor graph. We consider three 64-node acceptor graphs to be solved and three 6-node donor graphs, see Fig.~\ref{fig:example}. Table~\ref{tab:graphs} contains the details of the donor and acceptor graphs, including the total number of edges, the list of all subgraph components (labeled using the notations from Fig.~\ref{fig:heatmap2}), optimized $\gamma$ and $\beta$ parameters and the corresponding energy, energy of the optimal solution, and the approximation ratio. Graphs \#\#1--3, 5, and 6 consist of nodes with degrees 3 and 5, while the graph \#4 has nodes with degrees 1, 3, and 5. The optimized QAOA parameters for the donor and acceptor graphs were found by performing numerical optimization with 20 restarts, 200 iterations each. Table~\ref{tab:transfer} shows the results of the corresponding transfer of optimized parameters from the donor graphs \#\#1--3 to the acceptor graphs \#\#4--6, correspondingly. The approximation ratios obtained as a results of the parameter transfer in all three cases show only a 1--2\% decrease, compared to the ones obtained by optimizing the QAOA parameters for the corresponding acceptor graphs directly. These examples demonstrates the power of the approach introduced in this paper.

\begin{table*}
\begin{center}
\caption{}
    \begin{tabular}{||c c c c c c c c c||} 
    \hline
    Graph & Nodes & Edges & Subgraphs & $\gamma$ & $\beta$ & Energy & Energy (opt) & Approx. ratio\\ [0.5ex] 
    \hline\hline
    \#1 & 6 & 10 & (17, 24) & 2.22762 & 0.31316 & 6.26729 & 7.0 & 0.89533\\ 
    \hline
    \#2 & 6 & 11 & (18, 24, 44) & 2.24098 & 1.861 & 6.67106 & 8.0 & 0.83388\\
    \hline
    \#3 & 6 & 12 & (24, 44) & 2.24677 & 0.28448 & 7.18433 & 9.0 & 0.79826\\
    \hline
    \#4 & 64 & 102 & (2, 4, 16, 17, 22, 23, 40, 41) & 1.70967 & 1.9566 & 67.81171 & 89.0 & 0.76193 \\
    \hline
    \#5 & 64 & 144 & (8, 12, 22, 23, 40, 41, 42) & 2.47656 & 2.7577 & 92.76621 & 122.0 & 0.76038 \\
    \hline
    \#6 & 64 & 128 & (8, 12, 16, 17, 22, 23, 40, 41) & 2.48427 & 1.18309 & 83.85389 & 111.0 & 0.75544 \\ 
    \hline
    \end{tabular}

\label{tab:graphs}
\end{center}
\end{table*}

\begin{table}
\begin{center}
\caption{}
    \begin{tabular}{||c c c||} 
    \hline
    Transfer & Energy & Approx. ratio\\ [0.5ex] 
    \hline\hline
    \#1 $\to$ \#4 & 66.99604 & 0.75276 (-0.8\%)\\ 
    \hline
    \#2 $\to$ \#5 & 90.90949 & 0.74516 (-2.0\%)\\
    \hline
    \#3 $\to$ \#6 & 82.12825 & 0.73989 (-2.1\%)\\
    \hline
    \end{tabular}

\label{tab:transfer}
\end{center}
\end{table}

\section{Conclusions and outlook}\label{sec:conclusions}

Finding optimal QAOA parameters is a critical step in solving combinatorial optimization problems by using the QAOA approach. Several existing techniques to accelerate the parameter search are based on the advanced optimization and machine learning strategies. However, in most works, various types of global optimizers are employed. Such a straightforward approach is highly inefficient for exploration due to complex energy landscapes for hard optimization instances.

An alternative effective technique presented in this paper is based on two intuitive observations, namely, (a) the energy landscapes of small subgraphs exhibit ``well defined'' areas of extrema that are not anticipated to be an obstacle for optimization solvers (see Fig.~\ref{fig:landscapes}), and (b) structurally different sub-graphs may have similar energy landscapes and optimal parameters. A combination of these observations is important because in the QAOA approach, the cost is calculated by summing the contributions at the sub-graph level, where the size of a subgraph depends on the circuit depth $p$.

With this in mind, the overarching idea of our approach is solving the QAOA parameterization problem for large graphs by optimizing parameterization for much smaller graphs and reusing it. We started with studying the transferability of parameters between all subgraphs of random graphs with the maximum degree of 8. Good transferability of parameters was observed among even-regular and odd-regular subgraphs. In the same time, poor transferability was detected between even- and odd-regular pairs of graphs, in both directions, as shown in Figs.~\ref{fig:heatmap1} and~\ref{fig:heatmap2}. This experimentally confirms the proposed approach.

A remarkable demonstration on non-regular random graphs that generalizes the proposed approach is the transferability of the parameters from 6-node random graphs (at the sub-graph level) to 64-node random graphs, as shown in Fig.~\ref{fig:example}. The approximation ratio loss of only 1--2\% was observed in all three cases.

One may notice that we studied parameter transferability only for $p = 1$, where the subgraphs are small and transferability is straightforward. However, our preliminary work suggests that this technique will also work for larger $p$ which will require advanced subgraph exploration algorithms and will be addressed in the following work. Another promising future research direction is a generalization of parameter transferability for even-to-even and odd-to-odd degree graphs of different sizes.

This work was enabled by the very fast and efficient tensor network simulator QTensor developed at Argonne National Laboratory~\cite{qtensor}. Unlike state vector simulators, QTensor can perform energy calculations for most instances with $p \leq 3$, $d \leq 6$, and graphs with $N \sim\!1,000$ nodes very quickly, usually within seconds. For the purpose of this work, we computed QAOA energy for 64-node graphs with $d \leq 5$ at $p = 1$, which took a fraction of second per each execution on a personal computer. However, with state vector simulators, even such calculations would not have been possible due prohibitive memory requirements to store the state vector.

As a result of this work, finding optimized parameters for some QAOA instances will become quick and efficient, removing this major bottleneck in the QAOA approach and potentially removing the optimization step altogether in some cases, eliminating the variational nature of QAOA. Our method has important implications for implementing QAOA on relatively slow quantum devices, like neutral atom and trapped-ion hardware, for which finding optimal parameters may take a prohibitively long time. Thus, quantum devices will be used only to sample from the output QAOA state to get the final solution to the combinatorial optimization problem. Our work will ultimately bring QAOA one step closer to the realization of quantum advantage.

\section*{Acknowledgements}

A.G., D.L. X.L., I.S., and Y.A. are supported in part with funding from the Defense Advanced Research Projects Agency (DARPA). Y.A. is also supported in a part by the Exascale Computing Project (17-SC-20-SC), a joint project of the U.S. Department of Energy’s Office of Science and National Nuclear Security Administration, responsible for delivering a capable exascale ecosystem, including software, applications, and hardware technology, to support the nation’s exascale computing imperative. 
This work used in a part the resources of the Argonne Leadership Computing Facility, which is DOE Office of Science User Facility supported under Contract DE-AC02-06CH11357. The authors would like to thank Ruslan Shaydulin for insightful discussions.

\bibliographystyle{myIEEEtran}
\bibliography{ref}

\end{document}